# Holographic polymer-dispersed liquid crystal Bragg grating integrated inside a solid core photonic crystal fiber

Gianluigi Zito[*] and Stavros Pissadakis

*Foundation of Research and Technology-Hellas (FORTH), Institute of Electronic Structure and Laser (IESL), P.O. Box 1315, 71 110, Heraklion, Greece*
*Corresponding author: zito@iesl.forth.gr*

A polymer/liquid crystal-based fiber Bragg grating (PLC-FBG) is fabricated with visible two-beam holography by photo-induced modulation of a pre-polymer/LC solution infiltrated into the hollow channels of a solid core photonic crystal fiber (PCF). The fabrication process and effects related to the photonic bandgap guidance into the infiltrated PCF, and characterization of the PLC-FBG are discussed. Experimental data here presented, demonstrate that the liquid crystal inclusions of the PLC-FBG lead to high thermal and bending sensitivities. The microscopic behavior of the polymer/liquid crystal phase separation inside the PCF capillaries is examined using scanning electron microscopy, while further discussed.

A versatile platform for elaborating highly tunable and electric field-sensitive devices [1] has been demonstrated by infiltrating liquid crystals inside a photonic crystal fiber (PCF) [2] for inducing photonic bandgap (PBG) guidance [3-6]. In parallel, great effort has been invested into holographic polymer-dispersed liquid crystal (H-PDLC) technology, for the realization of tunable integrated optical elements, for sensor systems, head-up displays, optical interconnects, optical data storage [7,8] and lasing devices [9]. Such device development is based on the reorientation capabilities of the liquid crystal (LC) molecules under electrical field and the efficient opto-thermo-mechanical response of polymer/liquid crystal (P/LC) composites. P/LCs materials combined with holographic fabrication techniques, allow flexible processing for realizing volume diffraction phase gratings and photonic quasi-crystals [10,11]. Yet, Bragg grating fabrication into microstructured optical fibers [12] based on a composite material of polymer and liquid crystal has not been reported.

In this Letter, we report on the fabrication of a polymer/liquid crystal fiber Bragg grating (PLC-FBG) consisting of an H-PDLC phase grating recorded using two-beam holographic photo-polymerization of a suitable LC/photo-monomer solution (initially isotropic) infiltrated into the cladding channels of a solid core PCF. This hybrid H-PDLC PCF Bragg grating offers unique characteristics attributed to the infiltrated material, such as high sensitivity to bending. Moreover, analysis by scanning electron microscopy of the polymerized PDLC inside the PCF capillaries reveals phase-separation effects, characterized by the presence of LC nanodroplets embedded in the polymer matrix.

The photosensitive solution (PS) was infiltrated into the hollow channels (Ø ~ 2.85 μm) of the solid core PCF (LMA-10 manufactured by NKT Photonics Ltd.) by using a custom nitrogen pump apparatus providing an infiltration rate of 0.5 - 1.0 cm/min when operating at constant pressure of 3 - 6 bar. The solution optimized for such application allowed a compromise between low viscosity for efficient infiltration and high phase separation for effective diffraction and consisted of: dipentaerythrol-hydroxyl-penta-acrylate (DPHPA) monomer (40 - 45% w/w); N-vinylpyrrolidone (NVP) monomer (20 - 25% w/w) as cross-linking stabilizer and fluidity enhancer; rose bengal (RB) as photo-initiator (lower than 0.1 - 0.3% w/w to reduce the polymerization rate under exposure); N-phenylglycine (NPG) as co-initiator (0.1 - 0.5% w/w); and LC solution E7 to complete the mixture (typically ~ 35% w/w).

Fig. 1. Schematic representation of the experimental setup for

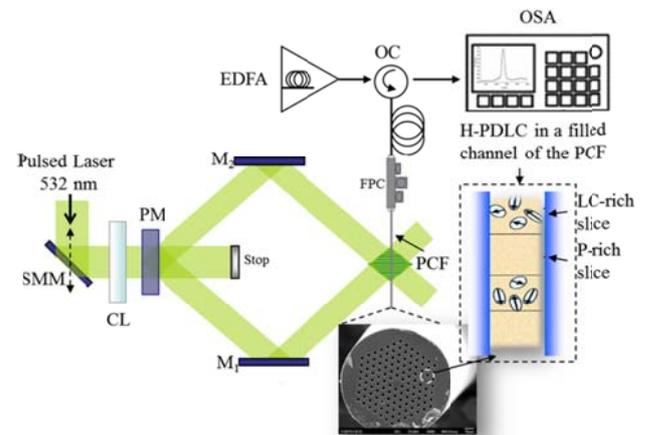

real-time inscription of the filled PCF.

The average refractive index of the isotropic PS, before polymerization, was in the range n ~ 1.52 - 1.54, measured at different relative ratios of NVP - DPHPA, with an Abbe refractometer. The nematic E7 liquid crystal has ordinary and extraordinary refractive indices $n_o = 1.52$ and $n_e = 1.73$ at 656 nm, respectively, and their temperature dependence may be found in [13].

Following infiltration, the high refractive index soft material inclusions turn the modified total internal reflection guidance into PBG guidance, reported before for other materials [13,14]. The PLC-FBG was inscribed directly into the filled fiber cladding by means of a two-beam interference holographic setup. A schematic representation of the experimental setup is shown in Fig. 1. The photo-polymerization was induced at room temperature by using a frequency doubled, 150 ps pulsed Nd:YAG laser ($\lambda_o$ = 532 nm) with pulse energy density $E_p$ = 5.0 mJ/cm$^2$ at a repetition rate of 5 - 10 Hz. A cylindrical lens (CL) was used to focus the beam at the writing plane. After spatial filtering, the beam had sizes 2.5 mm and 4.0 mm along the vertical and fiber directions, respectively. By using a phase mask (PM) with a pitch of 1064.7 nm and mirrors M1 and M2 to recombine the diffracted orders, an interference pattern with period ~ 539 nm was formed, corresponding to a measured Bragg wavelength $\lambda_B$ ~ 1558 nm on the PLC-FBG, tunable of tens of nanometers by varying the tilt angle of the mirrors M1 and M2. The grating was finally achieved by scanning the interference pattern along the fiber for a grating length of 5 - 8 mm by using a stepper motorized mirror (SMM).

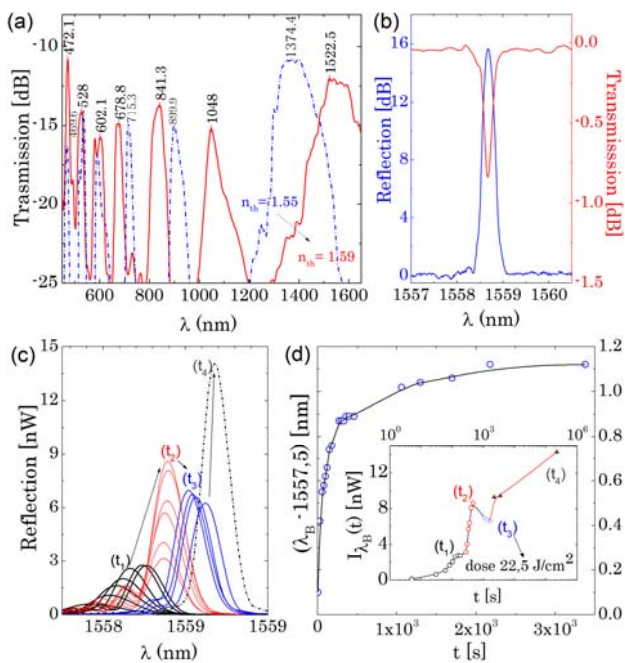

Fig. 2. (a) Transmission output of the infiltrated (dash blue line) and irradiated (solid red line) PCF. (b) Reflection and transmission spectral response of the PLC-FBG after exposure. (c) Evolution of the grating profile in reflection configuration during the grating formation. (d) Temporal behavior of the Bragg wavelength shift and peak amplitude (inset) during exposure of the grating.

The real-time grating formation and evolution is studied in reflection configuration. The grating formation was monitored in real-time by using, an erbium-doped fiber amplifier (EDFA) connected, through an optical circulator (OC), to a fiber patch cord (FPC) spliced to the infiltrated PCF. The time evolution of the probe signal reflected by the H-PDLC, while forming into the filled channel of the PCF, was measured by an optical spectrum analyzer (OSA). The exposure time t required for the grating finalization was about 10$^3$ s, corresponding to a cumulative irradiation dose of the order of 20 J/cm$^2$.

Prior exposure, the transmission spectrum of the filled PCF was measured independently with a Supercontinuum fiber laser (SuperK Compact, NKT Photonics Ltd., range of $\lambda$ ~ 350 - 2000 nm). The set-up used for this characterization was the same described in our preceding work [14]. This measurement revealed clear PBG transmission bands existing before and after exposure to ultraviolet radiation [Fig. 2(a)]. By cut-off wavelength calculation [14], we estimated the average refractive index of the H-PDLC inclusions to vary from $n_{th}$ ~ 1.55 (dash blue line) to ~ 1.59 (solid red line) upon polymerization.

In Fig. 2(b), an example of the reflection spectrum of the fabricated grating is shown (left scale); its strength in transmission is ~0.7dB. Fig. 2(c) represents the reflection spectra collected during the exposure process. In H-PDLCs, a phase separation is induced by the photo-polymerization process under laser recording of an interference pattern that turns an originally isotropic LC/photo-monomer mixture into a periodic modulated structure. Photo-polymerization of the monomer occurs in the bright regions of the writing pattern, initiating a diffusion process driving-out the LC-molecules towards the interference minima fringes. The morphology of the H-PDLC structure achieved depends strictly on the composition of the initial isotropic mixture and then, the photocuring parameters, i.e. the laser beam intensity and exposure time [8]. The time evolution of the grating formation, shown in Fig. 2(c) and 2(d), indicates the existence of three distinct phases during the H-PDLC formation. At phase ($t_1$) the grating peak amplitude increases with a large redshift because of the onset of photo-polymerization of the monomer (black lines); then the grating continues growing ($t_2$), however with minor redshift, due to the increasing of phase-separation between polymer and LC and consequent dielectric contrast augmentation (red lines). By continuing exposure, and after a large dose of irradiation, a grating overmodulation regime starts ($t_3$), which is characterized by a decreasing of the amplitude of the response (blue lines). When exposure is ceased ($t_4$), an increase of the grating amplitude (dash grey line) emerges with a simultaneous redshift progressing for several hours during post-curing at ambient condition (laser off), that is probably due to the stabilization of the grating morphology. The evolution of grating diffraction $I(\lambda_B)$ is compatible with the typical model of grating formation in PDLCs (see, as for instance, ref. [15] for an example of three-stage formation and [8] for the general model of the PDLC growth). In Fig. 2(d), the correspondent shift of the Bragg wavelength $\lambda_B$ versus the exposure time t is shown. A shift of the order of 1 nm was calculated to correspond to an effective refractive index variation of $\delta n_{eff}$ ~ 10$^{-3}$. Depending on the ratio of NVP-DPHPA in the pre-polymer solution and the energy density of the writing beam, a shift as high as 7.0 nm has been also measured (not shown here). The inset of the same figure shows the amplitude $I(\lambda_B)$ of the grating reflection as a function of time. The experimental points indicated with symbol (▲) refer to post-curing growth with laser off, i.e. following the irradiation dose of 22.5 J/cm$^2$ (reached after t ~ 2×10$^3$ s).

The PDLC infused gratings were also characterized for their temperature and bending sensitivity. Temperature characterization of the PLC-FBG shows a peak amplitude modulation and redshift that can be related both to the spectral shifting with temperature of the PCF transmission output and to the stabilization of the average refractive index of the grating. Fig. 3(a) shows the temperature dependence of the PLC-FBG reflection response characteristics measured within the nematic temperature range of the E7. As the operating temperature T (controlled by a Peltier plate) is increased towards the nematic-isotropic transition temperature $T_{NI}$ = 58 °C [13], the Bragg wavelength $\lambda_B$ redshifts and the corresponding peak amplitude $I(\lambda_B)$ is strongly attenuated.

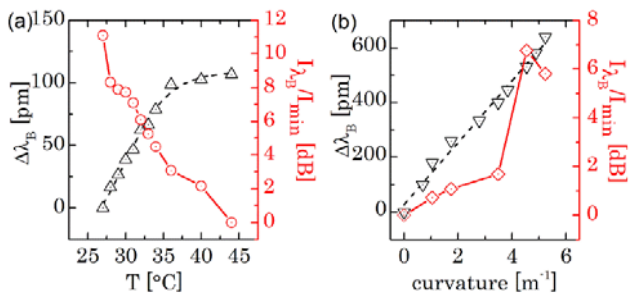

Fig. 3. (a) Bragg shift (left scale) and amplitude variation of Bragg peak (right scale) versus temperature of the PLC-FBG. (b) Bragg shift (left scale) and amplitude variation (right scale) versus curvature of the PLC-FBG. The amplitude of the Bragg peak is normalized to the minimum value $I_{min}$ measured in that range of curvature. Dash lines correspond to Boltzmann sigmoidal fit in (a) and linear fit in (b).

The red-shift effect can be partially explained by an increase of the grating period due to the thermal expansion of the periodic structure (and cladding). On the contrary, the reduced diffraction efficiency is likely due to the decrease of the average refractive index of the LC droplets with temperature, thereby resulting in a lower dielectric contrast between polymer and liquid crystals and a decrease of the average effective index. This last alone would lead to a blue-shift of $\lambda_B$. In fact, the saturated shift of the grating Bragg wavelength with temperature response with temperature suggests a competition between these two counterpart effects. In addition, it must be noted that the spectral position of the Bragg peak must coincide with a transmission band of the bandgap fiber to allow effective functioning of the grating. As a result, the variation of the fiber transmission output with temperature affects the diffraction efficiency by changing the evanescent coupling condition between the core and the in-cladding grating and can take part to the observed temperature behavior as well.

More importantly this "soft" PCF grating exhibited sensitivity to fiber bending. The measurements were carried out by undergoing the fiber at different curvatures near the grating position, at 25 °C, in reflection configuration. A significant relative red-shift of the Bragg peak, i.e. 110 ± 4 pm/m$^{-1}$, was measured, as shown in Fig. 3(b), while the relative amplitude of the peak increased. Fundamentally, the bending of the fiber leads to increased overlap with the high index inclusions on which the PBG guidance relies. In other words, there is larger overlap of the guiding mode in the bended PBG fiber regime with the periodic perturbation formed due to laser polymerization into the infiltrated strands. A tensile stress inducing a deformation of the average orientation of the LC-droplets may also result in an increase of the dielectric contrast of the grating; however, we cannot further confirm this.

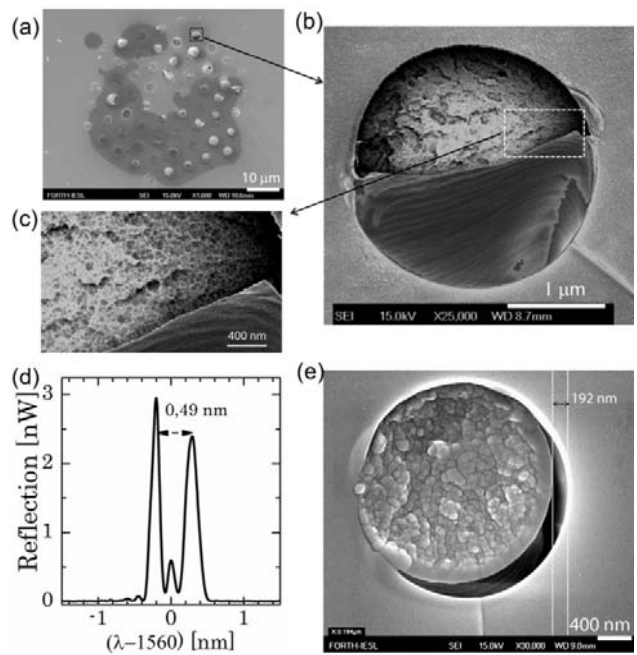

Fig. 4. (a) FE-SEM micrograph of the cleaved end-face of the infiltrated PCF. (b) Magnification area of micrograph (a): the grating inscribed is clearly visible into this channel of the PCF. (c) Detail of (b) showing an LC-nanodroplet structure. (d) Birefringent response of the PLC-FBG obtained under higher $E_p$ condition. (e) Magnification of the FE-SEM scan into the filled channel of the fiber, showing larger LC-droplets embedded in the polymer matrix, associated to the profile in (d).

Finally, we focused on the investigation of the laser photo-polymerization and phase separation processes taking place inside the PCF capillaries, where LC alignment issues can become prominent. By changing the energy density of the exposure, we have found that the grating microscopic morphology and spectral response were critically affected. In fact, too fast polymerization, related to higher energy densities, led to the formation of a birefringent grating response having a double-peak profile. We studied the microscopic morphology of the fabricated PLC-FBG by means of field emission-scanning electron microscopy (FE-SEM).

In Fig. 4(a), a FE-SEM micrograph of the PCF end-face, suitably cleaved at the grating position is presented. The micrograph of Fig. 4(b) is a magnification of one half-filled channel of Fig. 4(a) and shows the grating inscribed with $E_p = 5.0$ mJ/cm$^2$. The modulation of the infiltrated material is evident from the relief along the axial direction, i.e. the grating vector. A nanodroplet spatial structure embedded in the polymer matrix can be observed from the top layer of the transverse section of the grating, with LC-droplets in the range of 20 - 50 nm, as evident from Fig. 4(c). In order to emphasize the droplet matrix, an inverted contrast filter (MathWorks Matlab) is applied to the micrographs of Fig. 4(b) and (c). Such morphology, characterized by LC-nanodroplets of 25 nm on average size, was associated to the non-birefringent grating response, already reported in Fig. 2(b) and 2(c). Conversely, by irradiating the sample with higher energy density, i.e. $E_p = 7.7$ mJ/cm$^2$, a birefringent response was measured [Fig. 3(d)] and by similar FE-SEM analysis the correspondent morphological structure resulted to be characterized by larger LC-droplets having size of ~ 100 - 200 nm, as shown in Fig. 3(e). Moreover, it is worth to be noted that, as apparent from Fig. 4(e), a shrinkage effect of the polymer matrix was observed for the higher energy dose, but such shrinkage involved only very few capillaries on the external ring of the cladding without affecting the propagation. The detuning between the peaks typically observed for that birefringent profile was ≈ 0.5 nm. This result can be explained by considering an average orientation of the nematic LC-droplets with a small preferential alignment that induces a refractive index anisotropy $\delta n_{eff} \approx 5 \times 10^{-4}$ in the effective index of the guided mode. The possibility to determine the birefringent character of the PLC-FBG by changing the fabrication parameters may be significant to finalize the specific application of a potential device [16].

In conclusion, we reported on the fabrication of an H-PDLC Bragg grating, directly inscribed inside the cladding of a PCF, and experimentally characterizing its thermal and bending properties. This work aims at the development of a new platform of integrated fiber systems by merging the well-consolidated technology of soft matter polymer-liquid crystal composites with the unique guiding properties of photonic crystal fibers. The fabrication flexibility implied by a visible curing process may lead to the implementation of more complex geometries with single-step holographic techniques [11, 17]. Recent research showing in-fiber LC-tunability by electric field [1] as well as tuning potential offered by polymer/silica hybrid PCFs [18] suggests the possibility of further expanding the employment of LC/polymer composite for novel in-fiber applications. Research work is in progress for optimizing the diffraction efficiency, tuning the morphology and characterizing the electrical and magnetic response of the PLC-FBG. At same time, new possibilities are also being explored by introducing active inclusions and magnetic nanoparticles [19] in the liquid crystals.

**References**

1. S. Mathews, G. Farrell, Y. Semenova IEEE Photonics Tech. Lett. **23,** 408 (2011).
2. P. Russell, Science **299**, 358 (2003).
3. T. T. Larsen, A. Bjarklev, D. S. Hermann, J. Broeng, Opt. Exp. **11**, 2592 (2003).
4. T. T. Alkeskjold, L. Scolari, D. Noordegraaf, J. Lægsgaard, J. Weirich, L. Wei, G. Tartarini, P. Bassi, S. Gauza, S.-T. Wu, A. Bjarklev, Opt. Quant. Electron. **39**, 1009 (2007).
5. F. Du, Y.-Q. Lu, and S.-T. Wu, Appl. Phys. Lett. **85**, 2181 (2004).
6. T. R. Wolinski, K. Szaniawska, S. Ertman, P. Lesiak, A. W. Domanski, R. Dabrowski, E. Nowinowski-Kruszelnicki, and J. Wojcik, Meas. Sci. Technol. **17**, 985 (2006).
7. T. J. Bunning, L. V. Natarajan, V. P. Tondiglia, R. L. Sutherland, Ann. Rev. Mater. Sci. **30**, 83 (2000).
8. Y. J. Liu and X. W. Sun, Adv. in OptoElectron. Vol. **2008**, Article ID 684349.
9. G. S. He, T.-C. Lin, V. K. S. Hsiao, A. N. Cartwright, P. N. Prasad, L. V. Natarajan, V. P. Tondiglia, R. Jakubiak, R. A. Vaia, and T. J. Bunning, Appl. Phys. Lett. **83**, 2733 (2003).
10. G. Zito, B. Piccirillo, E. Santamato, A. Marino, V. Tkachenko, and G. Abbate, Mol. Cryst. Liq. Cryst. **465**, 371 (2007).
11. G. Zito, B. Piccirillo, E. Santamato, A. Marino, V. Tkachenko, and G. Abbate, Opt. Exp. **16**, 5164 (2008).
12. A. Cusano, D. Paladino, and A. Iadicicco, J. Light. Technol. **27**, 1663 (2009).
13. T. T. Alkeskjold J. Lægsgaard, A. Bjarklev, D. S Hermann, Anawati, J. Li, S.-T. Wu, Opt. Exp. **12**, 5857 (2004).
14. I. Konidakis, G. Zito, and S. Pissadakis, Opt. Lett. **37**, 2499 (2012).
15. G. Abbate, F. Vita, A. Marino, V. Tkachenko, S. Slussarenko, O. Sakhno, and J. Stumpe, Mol. Cryst. Liq. Cryst. **453**, 1 (2006).
16. D. C. Zografopoulos, E. E. Kriezis, and T. D. Tsiboukis, J. Lightwave Technol. **24**, 3427 (2006).
17. M. Infusino, A. De Luca, V. Barna, R. Caputo, and C. Umeton, Opt. Exp. **20**, 23138 (2012).
18. C. Markos, K. Vlachos, G. Kakarantzas, Opt. Mater. Exp. **2**, 929 (2012).
19. A. Candiani, W. Margulis, C. Sterner, M. Konstantaki, and S. Pissadakis, Opt. Lett. **36**, 2548 (2011).

"Full Citations"

1. S. Mathews, G. Farrell, Y. Semenova, Directional Electric Field Sensitivity of a Liquid Crystal Infiltrated Photonic Crystal Fiber, IEEE Photonics Tech. Lett. **23,** 408 – 410 (2011).
2. P. Russell, Photonic Crystal Fibers, Science 299, 358 (2003).
3. T. T. Larsen, A. Bjarklev, D. S. Hermann, J. Broeng, Optical devices based on liquid crystal photonic bandgap fibres, Opt. Exp. 11, 2592 (2003).
4. T. T. Alkeskjold, L. Scolari, D. Noordegraaf, J. Lægsgaard, J. Weirich, L. Wei, G. Tartarini, P. Bassi, S. Gauza, S-T.Wu, A. Bjarklev, "Integrating liquid crystal based optical devices in photonic crystal fibers," Opt. Quant. Electron. 39, 1009–1019 (2007).
5. F. Du, Y.-Q. Lu, and S.-T. Wu, Electrically tunable liquid-crystal photonic crystal fiber, Appl. Phys. Lett. 85, 2181 (2004).
6. T. R. Wolinski, K. Szaniawska, S. Ertman, P. Lesiak, A. W. Domanski, R. Dabrowski, E. Nowinowski-Kruszelnicki, and J. Wojcik, Meas. Sci. Technol. 17, 985 (2006).
7. T. J. Bunning, L. V. Natarajan, V. P. Tondiglia, R. L. Sutherland, Ann. Rev. Mater. Sci. 30, 83 (2000).
8. Y. J. Liu and X. W. Sun, Advances in OptoElectronics Vol. 2008, Article ID 684349.
9. G. S. He, T.-C. Lin, V. K. S. Hsiao, A. N. Cartwright, P. N. Prasad, L. V. Natarajan, V. P. Tondiglia, R. Jakubiak, R. A. Vaia, and T. J. Bunning, Tunable two-photon pumped lasing using a holographic polymer-dispersed liquid-crystal grating as a distributed feedback element, Appl. Phys. Lett. 83, 2733 (2003).
10. G. Zito, B. Piccirillo, E. Santamato, A. Marino, V. Tkachenko, and G. Abbate, "Computer-generated holographic gratings in soft matter," *Mol. Cryst. Liq. Cryst.*, vol. 465, pp. 371-378 (2007).
11. G. Zito, B. Piccirillo, E. Santamato, A. Marino, V. Tkachenko, and G. Abbate, Two-dimensional photonic quasicrystals by single beam computer-generated holography, Opt. Exp. 16, 5164-5170 (2008).
12. A. Cusano, D. Paladino, and A. Iadicicco, Microstructured Fiber Bragg Gratings, J. Light. Technol. **27**, 1663 - 1697 (2009).
13. T. T. Alkeskjold J. Lægsgaard, A. Bjarklev, D. S Hermann, Anawati, J. Li, S.-T. Wu, All-optical modulation in dye-doped nematic liquid crystal photonic bandgap fibers, Opt. Exp. 12, 5857 (2004).
14. Konidakis, G. Zito, and S. Pissadakis, Photosensitive, all-glass AgPO3/silicaphotonic bandgap fiber, Opt. Lett. 37, 2499 (2012).
15. G. Abbate, F. Vita, A. Marino, V. Tkachenko, S. Slussarenko, O. Sakhno, and J. Stumpe, New Generation of Holographic Gratings Based on Polymer-LC Composites: POLICRYPS and POLIPHEM, Mol. Cryst. Liq. Cryst., 453, 1 (2006).
16. Dimitrios C. Zografopoulos, Emmanouil E. Kriezis, and Theodoros D. Tsiboukis, "Tunable Highly Birefringent Bandgap-Guiding Liquid-Crystal Microstructured Fibers," J. Lightwave Technol. **24**, 3427- (2006)
17. M. Infusino, A. De Luca, V. Barna, R. Caputo, and C. Umeton, Periodic and aperiodic liquid crystal-polymer composite structure realized via spatial light modulator direct holography, Opt. Exp. 20, 23138 (2012).
18. C. Markos, K. Vlachos, G. Kakarantzas, Guiding and thermal properties of a hybrid polymer-infused photonic crystal fiber, Opt. Mater. Exp. 2, 929 - 941 (2012).
19. A. Candiani, W. Margulis, C. Sterner, M. Konstantaki, and S. Pissadakis, Phase-defected Bragg gratings realized in microstructured optical fibres utilizing infiltrated ferrofluids, Opt. Lett. 36, pp. 2548-2550 (2011).